# Evolution of cavities in BCC-Fe with coexisting H and He under fusion environments

Jin Wang[a], Fengping Luo[a], Tao Zheng[a], Bowen Zhang[a], Yuxin Liu[a],

Denghuang Chen[a], Xinyue Xie[a, b], Mohan Chen[a, b], Hong-Bo Zhou[c],

Fei Gao[d], Jianming Xue[a], Yugang Wang[a, *], and Chenxu Wang[a, *]

[a] *State Key Laboratory of Nuclear Physics and Technology, Center for Applied Physics and Technology, Peking University, Beijing, 100871, China*

[b] *HEDPS, College of Engineering and School of Physics, Peking University, Beijing 100871, China*

[c] *Department of Physics, Beihang University, Beijing, 100191, China*

[d] *Department of Nuclear Engineering and Radiological Science, University of Michigan, Ann Arbor, MI, 48109, USA*

* Corresponding authors. E-mail addresses: ygwang@pku.edu.cn (Yugang Wang). *and* cxwang@pku.edu.cn (Chenxu Wang).

## Abstract

In the fusion environment, understanding the synergistic effects of transmutation-produced hydrogen (H), helium (He), and irradiation-induced displacement damage in iron-based alloys is crucial for the development of structural materials for fusion reactors. When H and He atoms are simultaneously introduced into the matrix, the interaction between irradiation-induced cavity defects (voids and bubbles) with H and He, along with their evolutionary behavior remains poorly understood. In this study, the evolutionary behavior of cavities in body-centered cubic (BCC) iron (Fe) with H and He atoms is systematically investigated through a combination of molecular dynamics (MD) calculations and statistical thermodynamics. First, an efficient and suitable set of Fe-H-He ternary potential functions for describing interatomic interactions is established. Based on the newly developed MD model, the evolutionary behavior of H/He atoms and cavities is systematically investigated under various temperature and cavity structure conditions. Specifically, the kinetic process of H/He capture by cavities is elucidated for different scenarios. Additionally, thermodynamic analyses are employed to assess the feasibility of cavity trapping of H under varying conditions. The results exhibit strong consistency with experimental results and provide significant evidence supporting the formation of the core-shell structure (where He is confined at the cavity center while H accumulates at the surface) from both kinetic and thermodynamic perspectives. This work provides mechanistic insights into the nucleation and growth of cavities over extended temporal and spatial scales in the presence of H-He synergies.

## 1. Introduction

Fusion energy has emerged as a promising and sustainable energy source, potentially providing a substantial resolution to the ongoing global energy crisis. However, the advancement of nuclear structural materials capable of enduring extreme and demanding service conditions remains a key challenge in the development of next-generation nuclear energy systems [1–3]. Deuterium-tritium fusion reactions expose structural materials to extreme temperatures and a 14 MeV neutron spectrum, leading to significant atomic displacement damage and transmutation nuclear reactions. In addition to the formation of radiation-induced defects and defect clusters, high-energy nuclear transmutation reactions (such as (n, p) and (n, α) reactions) generate substantial amounts of low-solubility transmutation gases, primarily hydrogen (H) and helium (He) [2]. Due to the absence of high-flux fusion-neutron sources, triple-ion beams (heavy ion + H + He) are considered a promising alternative for studying the behavior of materials under fusion-neutron irradiation. Most triple-beam ion irradiation results demonstrate that strong synergistic interactions among H, He, and radiation-induced displacement damage significantly influence cavity swelling [4–11] and tritium retention [12–14] in various candidate structural materials for fusion reactors.

Reduced Activation Ferritic/Martensitic (RAFM) steels stand out as leading candidates for use in demonstration fusion nuclear power plants. Their low activation properties and body-centered cubic (BCC) structure provide inherent resistance to irradiation-induced swelling [15–17]. In recent decades, numerous notable studies have examined the effectiveness of RAFM steels in fusion applications within the context of the synergies between radiation damage and transmutation gas production. Wakai et al. [5,6] demonstrated significant swelling enhancement in F82H steel under triple-beam ($Fe^{2+}$, $He^+$, and $H^+$ ions) and dual-beam ($Fe^{3+}$ and $He^+$ ions) irradiation at 470-600°C to 50 dpa, revealing strong H-He synergistic effects. Tanaka et al. [7] subsequently demonstrated a 10× swelling increase in Fe-12Cr under triple-beam irradiation (510°C, 50 dpa, 40/10 appm/dpa H/He) versus dual-beam conditions, attributed to increased cavity size despite 50% lower cavity density. Recent studies on RAFM steels demonstrate that H/He co-implantation enhances cavity swelling compared to single/dual-ion irradiation [10,11,18]. Zimber et al. [18] found larger cavities in Eurofer97 under triple-beam irradiation, while Clowers et al. [10] observed increased cavity density and size in three ferrite/martensitic alloys with H+He co-implantation. However, excessive H implantation in F82H suppressed swelling due to small bubble nucleation [11], highlighting the complex role of H in cavity evolution under varying temperatures, doses, and implantation rates. The precise role of H in cavity nucleation/growth, its quantitative impact, and the atomic-scale mechanisms driving H-cavity interactions that exacerbate swelling remain unresolved [11,19].

Moreover, the recent experimental findings [10,11,18] have provided key insights into the interaction of H and He atoms with cavities, as well as the formation of H-He-Vacancy (V) cavity configurations following the trapping of H and He by cavities in various RAFM steels. These results provided clear evidence for the H-He-V cavity configuration at an irradiation temperature of 450 °C (723 K) using electron energy loss spectroscopy (EELS). This analysis revealed that H forms a halo structure at the periphery of the cavity, while He is present inside the cavity. However, Myers et al. [20] measured the enthalpy of deuterium capture by small He bubbles as 0.75 ± 0.07 eV, suggesting H de-trapping at cavity surfaces occurs near 327 °C (600 K) [21]. It is widely accepted that above this de-trapping temperature, cavities cannot effectively trap H atoms from the BCC-Fe matrix. Nevertheless, the underlying physical mechanism responsible for this behavior in high-temperature irradiation experiments remains unclear.

Computational studies provide critical atomic-scale insights into H-He interactions in structural materials. Previous density-functional theory (DFT) calculations and molecular dynamics (MD) simulations [22–25] have revealed important interaction mechanisms between H, He, and vacancy defects in BCC-iron (Fe). Since both He and H exhibit a strong thermodynamic preference for low-electron-density regions such as vacancies, their intrinsic repulsive interaction is effectively screened by the stronger He-vacancy (He-V) and H-vacancy (H-V) binding. Consequently, vacancies serve as stable trapping sites that enable H-He co-segregation despite their mutual repulsion [22]. Furthermore, Hayward et al. [23] employed a hybrid MD and Monte Carlo (MC) approach to determine the lowest-energy configurations of H and He in cavities. Their results revealed a stable structure with a He-rich core surrounded by a shell of H atoms, where H is attracted to the cavity surface. This seems to provide a plausible explanation for the experimentally observed co-segregation of H and He [10,11,18]. However, since the study is based on ground-state energetics, it cannot account for the evolution of H-He-V cavities under actual high-temperature irradiation conditions. Critical factors such as thermal effects, dynamic defect interactions, and non-equilibrium processes—all of which dominate in real irradiation environments—are inherently absent in static energy calculations [19]. To accurately explore the dynamic coevolution of H and He within cavities under irradiation, kinetic simulations at elevated temperatures are essential.

Few MD studies have investigated the kinetic evolution of coexisting H and He atoms in BCC-Fe-containing cavities, primarily due to the absence of a reliable and computationally efficient Fe-H-He ternary interatomic potential. Hayward et.al [23] first developed a set of interatomic potentials to describe all interactions among H, He, and Fe. Their analysis of bubble energetics and structure revealed a synergetic mechanism: H enhances He-induced loop punching, promoting bubble growth.

However, the MD simulations employed the Fe-H potential developed by Ramasubramaniam et al. [26], which exhibits a strong attraction for H-H interactions, contrasts with the results of DFT calculations [27,28]. This significant deficiency leads to an overestimation of H concentration near defects (e.g., dislocation cores and crack tips) and may induce artificial H clustering in simulations. To resolve this limitation, Wen [28] developed a new many-body interatomic potential for H in BCC-Fe. This improved potential accurately describes both the behavior of individual H atoms in Fe and reliably captures H-H interactions. Additionally, Chen et al. [24] studied H trapping behavior in He bubbles using an inappropriate Lennard-Jones potential, reporting increased H accumulation with higher He/V ratios. However, this prediction conflicts with experimental measurements [29] and DFT calculations [22,25], both demonstrating that He suppresses H retention in bubbles. Recently, Wu et al. [30] developed a machine learning (ML)-based deep potential (DP) for Fe-H-He ternary systems. This potential enables accurate and quantitative prediction of interactions between small H/He clusters and vacancy defects. While the DP-FeHHe achieves high computational accuracy, its efficiency is nearly 100 times lower than that of empirical potentials [30]. However, the DP-Compress method may help to further increase the efficiency [31]. In addition, Huang et al. [32] developed a new Fe-H-He ternary empirical potential via systematic fitting and optimization, subsequently employing it to investigate the synergetic evolution of H and He with defects.

All the above experiments and calculations provide significant insights into the retention or trapping of H in bubbles under various conditions. Unfortunately, when H and He atoms coexist in the material, research on the interaction of irradiation-induced cavity defects (voids and bubbles) with H and He, and the evolutionary behavior remains an area of limited comprehension. The evolution of H-He-V cavities is a dynamic process, involving both the formation and dissolution of these complexes. The trapping and de-trapping of H and He are key factors driving this evolution. Therefore, it is essential to consider the dynamic evolutionary behavior between the cavity and the H and He atoms under various cavity structures, H/He concentrations, and different temperature conditions. Furthermore, a critical gap persists in understanding how cavities in BCC-Fe with co-existing H and He atoms form stable core-shell configurations across different temperature regimes.

This study combines MD simulations with statistical thermodynamics to investigate the dynamic evolutionary behavior between cavities and H/He atoms, and the thermodynamic feasibility of H capture by cavities under different conditions (including temperature, H concentration, and cavity size). Based on the newly established MD model, it is obtained that higher temperatures drive interstitial H escape but favor $H_2$ molecule formation in cavities, reflecting a balance between H escape and cavity-mediated storage. Moreover, high He/V ratios limit $H_2$ storage capacity but are partially offset by He-induced cavity expansion, increasing surface trapping sites. At high He/V ratios, extended strain fields broaden H-trapping regions. In all studied cases, the cavity surface acts as the energetically favorable trapping site for H atoms, resulting in a stable "shell-like" structure within the cavity. Critically, the thermodynamic calculations provide substantial support for the kinetic evolution results. Investigations into the evolutionary mechanisms of cavities in BCC-Fe with coexisting H and He can provide crucial insights into the long-term behavior of H-He-V cavities, ultimately enabling better prediction and control of their microstructural evolution.

## 2. Calculational methodology

### 2.1. Interatomic Potentials

This study employed a newly developed hybrid-style Fe-H-He ternary potential and was performed using the Large-scale Atomic/Molecular Massively Parallel Simulator (LAMMPS) [33]. The construction of the Fe-H-He interatomic potential requires six interaction parameter sets: Fe-Fe, Fe-H, Fe-He, H-H, H-He, and He-He. In this work, the potential function is developed following the methodology established in the previous study [32]. For the Fe-He system, the "s-band model" potential developed by Gao et al. [34] was selected to describe the many-body interactions contributed by electrons. This interatomic potential employs the Ackland-Mendelev potential [35,36] to describe the Fe-Fe pair and many-body interactions, and the Aziz potential [37] for the pair potential utilized in He-He interactions, respectively. This set of potentials performs better in estimating the overall binding energy than the previous Fe-He potentials [38]. Moreover, the potential has been successfully used to evaluate the temperature dependence of $He_nV_m$ clusters formed by self-trapping [39] and the effect of He clusters on grain boundary migration [40,41] in BCC-Fe. The empirical embedded atom method (EAM) potential developed by Wen [28] has been selected for the Fe-H system, exhibiting a high degree of concordance with the DFT results [30]. It provides an accurate refinement of the H-H interactions, enabling an accurate description of the bond strength and length of the $H_2$ molecule. Furthermore, the EAM potential has been successfully employed in describing H interactions with vacancies, dislocations, and grain boundaries [42–44].

Although the van der Waals interactions between H and He atoms are relatively weak, it is of great importance to select an accurate interaction potential between these two elements to facilitate the investigation of H/He clusters and their interactions with vacancies. The conventional Lennard-Jones (L-J) pair potential is unable to provide a theoretically plausible form of repulsion at short-range distances. To correct this, the Ziegler-Biersack-Littmark (ZBL) potential is applied for close-range interactions. For this study, we adopted the H-He interaction potential developed by Hayward et al. [23]. The final interatomic potential was derived by fitting the functional forms of these potentials along with their associated parameters.

## 2.2. Physical parameter calculations

For a reliable study of the trapping and de-trapping behavior of H/He atoms in the presence of defects, it is essential to ensure the accuracy of the atomic stress calculations within the system. This is a crucial component in the subsequent analysis of local stress field variations and the calculation of bubble pressure. According to the Virial definition [45], the per-atom stress in the system is defined as

$$\sigma_{\alpha\beta}(i) = -\frac{1}{V_i}\left(m_i v_i^\alpha v_i^\beta + \frac{1}{2}\sum_j F_{ij}^\alpha r_i^\beta\right) \qquad (1)$$

where the term $V_i$ denotes atomic volume, $m_i$ represents atomic mass, and $v_i$ signifies the thermal vibration velocity of atoms. $F_{ij}$ refers to the interaction force between atoms $i$ and $j$. $\alpha$ and $\beta$ represent the Cartesian components. The initial term of Eq. (1) indicates the thermal kinetic energy, while the final term represents the virial resulting from the pairwise interaction between atom $i$, situated at position $r_i$, and atom $j$ at position $r_j$. The volume of each atom was estimated by employing the Voronoi tessellation method in conjunction with the VORO++ package [46].

Given that stress is a tensor quantity, its overall magnitude can be evaluated by combining all its matrix elements, which are described by a symmetric 3×3 matrix $\sigma_{\alpha\beta}(i)$, involving six independent matrix elements [47]. A suitable scalar is known as the von Mises stress, which is denoted as

$$\sigma_{vm} = \sqrt{\frac{1}{2}[(\sigma_{11} - \sigma_{22})^2 + (\sigma_{22} - \sigma_{33})^2 + (\sigma_{33} - \sigma_{11})^2] + 3(\sigma_{12}^2 + \sigma_{23}^2 + \sigma_{31}^2)} \quad (2)$$

It is noteworthy that the change in the stress field of neighboring atoms resulting from defects can be characterized by the per-atom von Mises.

Additionally, the average pressure [48,49] within a bubble (including the contribution from H/He atoms) is derived via

$$P = -\frac{1}{3V}\sum_i^n[\sigma_{11}(i) + \sigma_{22}(i) + \sigma_{33}(i)] \quad (3)$$

where $\sigma_{11}(i)$, $\sigma_{22}(i)$, and $\sigma_{33}(i)$ are the diagonal components of the stress tensor for the $i^{th}$ H/He atom, and $V$ represents the total Voronoi volume of the bubble.

### *2.3. Simulation details*

The flowchart illustrating the simulation process and the schematic diagram of the model construction are presented in Fig. 1. The perfect alpha-Fe system was initially generated via a simulation box with dimensions of 50 $a_0$ × 50 $a_0$ × 50 $a_0$ along the $x$-, $y$-, and $z$-directions, containing $2.5 \times 10^5$ Fe atoms. Here, the $a_0$ represents the lattice constant of a perfect BCC-Fe (2.8553 Å). Based on experimental results [10,11,18], the cavities (both bubbles and voids) in RAFM steels have a broad size distribution, with diameters ranging from 2 ~ 3 to 20 ~ 30 nm at 573 ~ 723 K irradiation temperature. The peak cavity size distribution is observed to be approximately 4 ~ 6 nm. Moreover, the results of He implantation in martensitic steels revealed that the He/V ratio in their He bubbles ranged from 0.23 to 0.81, as determined by EELS [50,51]. Thus, the constructed cavity radius was selected in the present study to be 0.57 ~ 2.86 nm (65 ~ 8462 vacancies), and the He/V ratios of 0, 0.5, and 0.8 are chosen to investigate the evolutionary behavior of H and He for cavity defects. After constructing the irradiation-induced cavity, the conjugate gradient (CG) algorithm was used to minimize the system energy. The entire simulated system was initially relaxed using an isobaric-isothermal ensemble (NPT) to eliminate the influence of residual stress, followed by equilibration at the target temperature and zero pressure using a 1 fs timestep for at least 0.1 ns.

Subsequently, vacuum layers (thickness of 10 $a_0$) were introduced along the *x*-, *y*-, and *z*-directions in the atomic configuration containing H/He atoms and the cavity (void or bubble), as illustrated in Fig. 1(b). The existing system was minimized by deploying the CG energy minimization function, and the model was equilibrated through the implementation of a canonical ensemble (NVT) at least 0.2 ns. The application of the vacuum layer enables the escape of H/He atoms from the BCC-Fe matrix under thermodynamic effects. The cyclic insertion method was employed in this study, whereby H/He atoms were introduced at intervals of at least 200,000 time steps (~0.2 ns) per cycle, for a total of 100 cycles (~20 ns). Notably, the accuracy of the results and the computational efficiency have been balanced as much as possible at the spatial and temporal scales of the MD, despite the excessively high rate of introduction of H/He atoms in the substrate material. The present model effectively evaluates the evolutionary behavior between cavities and H/He atoms by avoiding the formation of substantial He clusters within the system and facilitating the escape of atoms at elevated temperatures.

In addition, Fe-based structural materials served in fusion environments produce H and He atoms in transmutation reactions at rates of about 40 appm/dpa and 10 appm/dpa, respectively [19,52]. Therefore, the ratio of inserted H and He atoms was maintained at 4:1 when the cavity was initially constructed as a void in this study. For the He bubble, H atoms were similarly introduced into the system, thereby facilitating observation of the dynamic evolution between H atoms and the He bubble. The evolutionary behavior of the cavity trapping H/He atoms were considered for H and He concentrations ranging from 0 to 16,000 and from 0 to 4,000 appm, respectively. Following the completion of the implantation process, the system was subjected to a NVT ensemble at the target temperature (300 ~ 973 K) to investigate the long-term dynamic evolution of H/He atoms and cavities. It is worth emphasizing that the application of the vacuum layer enabled the escape of H/He atoms during the dynamic evolution phase, despite the absence of further H/He atoms introduced. Besides, the snapshot visualizations under different atomic configurations were subjected to analysis and visualization using a combination of Wigner-Seitz analysis, Cluster analysis, Voronoi analysis, and Construct Surface Mesh (CSM) algorithm as implemented in OVITO [53].

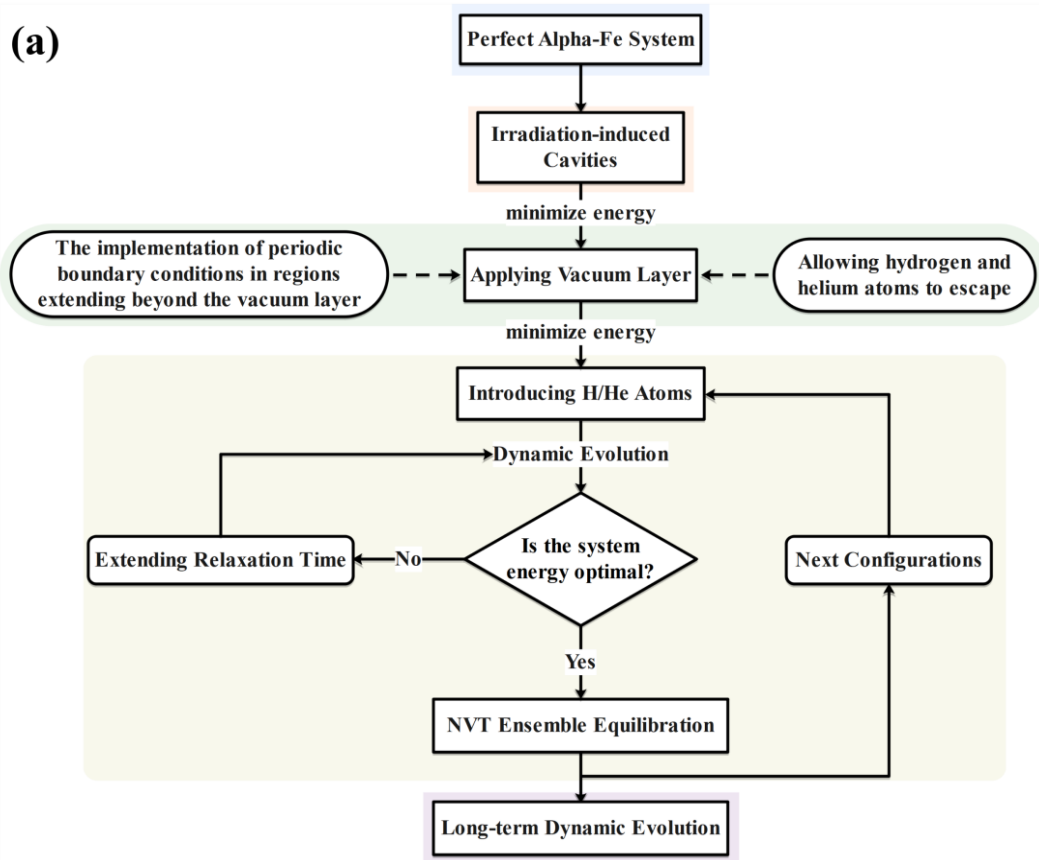

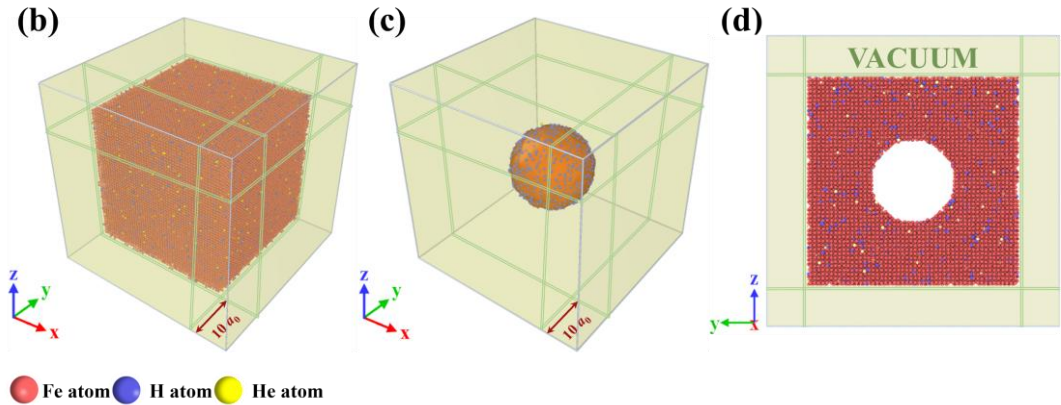

**Fig. 1.** Principal flowchart of the simulation process (a) and schematic diagram of the model construction (b-d). (b) The conformational diagram of H and He atoms in the matrix of BCC-Fe containing a cavity, and the region of 10 times the lattice constant beyond the atomic configuration is identified as the vacuum layer. (c) The constructed cavities are characterized by using Constructing Surface Mesh, with the surfaces of the cavity subsequently rendered in orange. (d) A central slice map of the atomic configuration depicted in Fig. 1(b) (with a thickness of 10 Å) is employed to observe the distribution of atoms within and outside the cavity.

## 3. Results

### *3.1 Temperature effects on the trapping of H/He*

#### *3.1.1 He forms a "core-like" structure within the cavity*

The morphology and location of the H and He atoms coexisting in the BCC-Fe matrix are influenced by many factors, including ambient temperature, H/He concentration, and cavity structure (different sizes and He/V ratios). Based on the newly developed MD model, the effects of the above factors on the evolutionary behavior of H/He atoms and cavities are investigated. It should be noted that the H/He concentrations inside and outside the cavity are defined as the number of H/He atoms divided by the cavity volume and the substrate volume (excluding the cavity), respectively.

The evolution of He concentration ($C_{He}$) inside and outside the same cavity (with a radius of 2.86 nm and a He/V ratio of 0) is analyzed at different temperatures, as illustrated in Figs. 2 (a-c). During the cyclic introduction of the H+He atoms stage, the vacuum layer allowed atoms to escape from the matrix, yet He atoms were persistently trapped by the cavity. Due to its low host metal electron density, the cavity acts as a strong trapping site for He [54], with a minimum binding energy of approximately 2.0 eV [22,23]. During the dynamic evolution phase, the $C_{He}$ within the cavity remains essentially constant despite the absence of further atom introduction. This stability arises from the high energy barrier [55] that prevents He escape once trapped, ensuring that at temperatures below this threshold, He atoms remain stably confined within the core of the cavity. However, during the cyclic introduction of H+He atoms, the number of He atoms entering the core increases with temperature. Concurrently, the $C_{He}$ inside and outside the cavity exhibited corresponding changes, specifically an increase inside and a decrease outside. As shown in Figs. 2 (a-c), the He capture efficiency of the cavity increases with temperature, reaching values of $1.15×10^8$ nm$^{-3}$s$^{-1}$, $2.28×10^8$ nm$^{-3}$s$^{-1}$, and $2.65×10^8$ nm$^{-3}$s$^{-1}$ at three respective temperatures. This increase arises from thermally enhanced He atoms diffusion, leading to intensified accumulation at defect sites. The behavior mirrors the temperature-driven formation and growth of He clusters and bubbles at grain boundaries [40,41].

The distribution of cumulative He density in the cavity at the three temperatures is presented in Fig. 2 (d). These data are obtained by time-averaging the configurations at all time steps during the dynamic evolution stage. The region at a radial distance of 28-32 Å from the cavity center corresponds to the outer surface of the cavity. The He atoms trapped by the cavity are distributed throughout the core, and the spatial distribution probability of He in the cavity remains constant independent of temperature. In a BCC-Fe matrix containing a cavity, the interstitial He atoms are either (i) trapped by the cavity, (ii) self-trapping in the matrix to form He clusters [23], or (iii) escape from the matrix. When H and He atoms coexist in the matrix, the He$_n$V$_m$ clusters formed by self-trapping further capture interstitial H atoms to form H$_i$He$_n$V$_m$ clusters. Fig. 2 (e) presents the evolution of the number and size of H$_i$He$_n$V$_m$ clusters at each temperature during the cyclic introduction of H+He atoms (within 20 ns). An increase in temperature results in a notable decline in the number and size of clusters outside the cavity. Furthermore, the correspondence between the size and number of clusters at the end of the H+He atoms cycle introduction is shown in Fig. 2 (f). The results demonstrate that increasing temperature elevates the kinetic energy of He atoms, thereby destabilizing the formation of small clusters and reducing the quantity and size of $H_iHe_nV_m$ clusters. Consequently, interstitial He atoms exhibit higher probabilities of either escaping the lattice or being trapped by cavities, with elevated temperatures promoting cavity-mediated He retention.

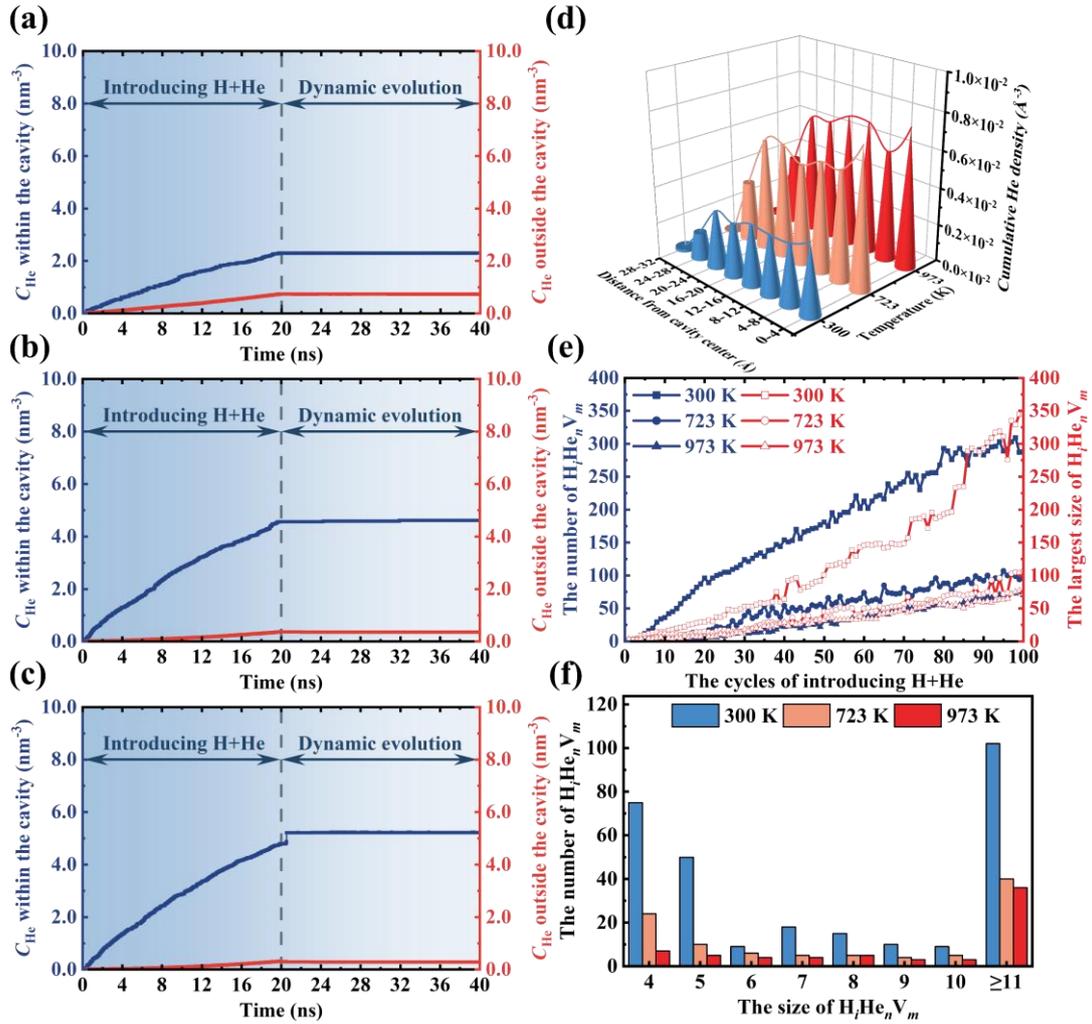

**Fig. 2.** (a-c) Evolution of He concentration ($C_{He}$) inside and outside the cavity at temperatures of (a) 300 K, (b) 723 K, and (c) 973 K during different stages. (d) The distribution of cumulative He density within the cavity under full-time averaging during the dynamic evolution stage. (e) The number and largest size of $H_iHe_nV_m$ clusters outside the cavity at each cycle. (f) The correlation between cluster size and number at the end of the cyclic introduction of H+He atoms.

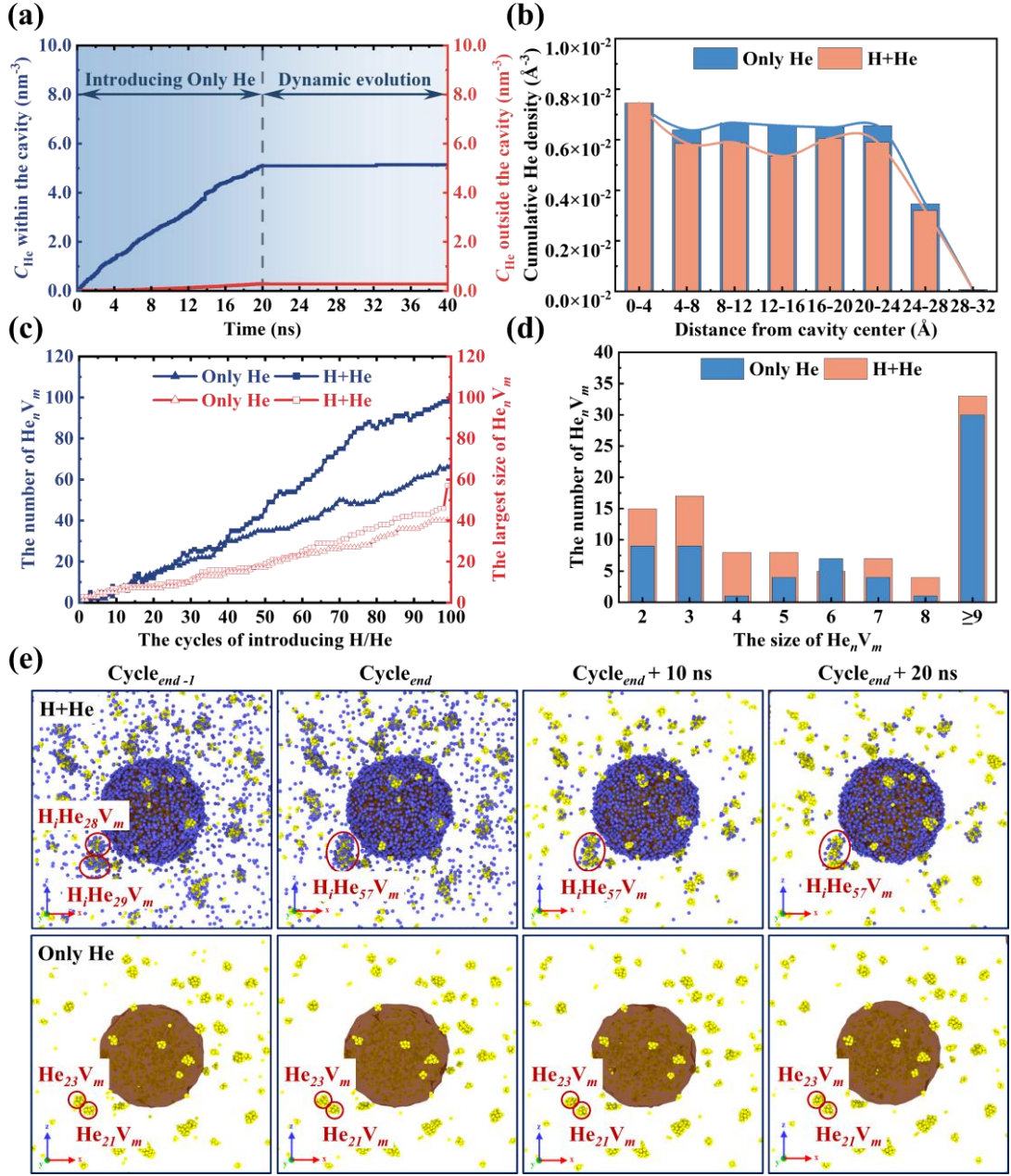

**Fig. 3.** (a) Evolution of He concentration ($C_{He}$) inside and outside the cavities when only He atoms are introduced at 723 K. (b) The distribution of cumulative He density within the cavities under full-time averaging during the dynamical evolution stage. (c) The number and largest size of $He_nV_m$ clusters at each cycle under conditions of co-introduction of H+He and introduction of He only. (d) The correlation between cluster size and number at the end of the cyclic introduction of H/He atoms. (e) Configuration diagrams of the cavity and the H/He atoms, viewed along the [101] direction. The H and He atoms are colored blue and yellow, respectively, and the cavity surface is highlighted orange using the CSM algorithm.

To further clarify the influence of H atoms in the BCC-Fe matrix on the behavior of void-captured He atoms, Figs. 3 (a) and (b) illustrate the variation of $C_{He}$ and the He atoms distribution averaged over the entire period when only He is introduced at 723 K. Comparative analysis of Fig. 3 (a) and Fig. 2 (b) reveals that the efficiency of He capture by the cavity increases when only He is introduced. In the absence of an interstitial H atom within the matrix, the concentration of He atoms trapped by the cavity is observed to increase from 4.56 nm$^{-3}$ to 5.14 nm$^{-3}$. However, since vacancy defects are occupied by H and He at different sites [23,56,57], the distribution of He atoms within the cavity remains unaltered by the presence of H, as shown in Fig. 3 (b).

Moreover, Fig. 3 (c) and (d) provide the evolution of the number and size of He$_n$V$_m$ clusters outside the cavity at each cycle, as well as the correlation between cluster size and number at Cycle$_{end}$. The presence of interstitial H atoms hinders the migration of interstitial He atoms, which in turn facilitates the self-trapping of He clusters and the subsequent formation of more H$_i$He$_n$V$_m$ clusters. Ultimately, it is unfavorable for the H and He atoms in the matrix to be trapped by the cavities. As evidenced in Fig. 3 (e), the H atoms exert a considerable influence on He$_n$V$_m$ clusters situated outside the cavity. The aggregation process of two smaller H$_i$He$_n$V$_m$ clusters to form a larger cluster (H$_i$He$_{28}$V$_m$ + H$_i$He$_{29}$V$_m$ → H$_i$He$_{57}$V$_m$) is mediated by H atoms. In contrast, two He$_n$V$_m$ clusters at the same position remain stable in the matrix throughout the evolution process in the absence of H atoms. Besides, the absorbed H atoms will then form a similar shielding layer on the surface of the cavity, which may act to prevent further He atoms from entering the cavity. Overall, when H and He atoms coexist in the BCC-Fe matrix, interstitial H enhances the self-trapping effect of He atoms, which partially suppresses He atoms accumulation in pre-existing cavities. Furthermore, irrespective of the presence of H in the BCC-Fe matrix, once He atoms are trapped by the cavity, a "core-like" structure is formed within the cavity.

### 3.1.2 H forms a "shell-like" structure within the cavity

The behavior of H in the metal matrix differs from that of He due to their distinct electronic structures [19]. In contrast to He in BCC-Fe, H atoms exhibit a low propensity for self-captured cluster formation in the unstressed matrix [57–59].

Furthermore, H atoms are difficult to move as interstitials in a perfect lattice over long periods. As a result, introduced H atoms either: (i) become trapped at cavities, (ii) are captured by $H_iHe_nV_m$ clusters outside cavities, or (iii) escape from the matrix. However, similar to He in the BCC-Fe matrix, H in metallic materials accumulates preferentially in regions of reduced electron density. The cavity surface is also an effective trapping region for H due to its lower electron density [25,54]. As demonstrated in Figs. 4 (a-c), the variation of H concentration ($C_H$) inside and outside the cavity with a radius of 2.86 nm and a He/V ratio of 0 at different temperatures is also investigated. Although H atoms exhibit a lower binding energy to vacancies (~0.75 eV) [56,60] compared to He atoms, the $C_H$ within cavities remains significantly higher than in the surrounding matrix across all simulated temperatures, confirming that cavities effectively trap and retain H atoms. With increasing temperature, interstitial H atoms either escape from the matrix or become trapped in cavities. Consequently, the $C_H$ outside cavities gradually decrease with rising temperature, while the $C_H$ inside cavities follows a nonlinear trend.

Moreover, Figs. 4 (d-f) display the fraction of H atoms within a distance of the cavity center at three temperatures at distinct times. The local magnification corresponds to the blue banded region for demonstrating the H fraction at the inner and outer surfaces of the cavity. Note that the "Initial" time point in these figures corresponds to the first introduction of H+He atoms into the system. At the initial introduction of atoms, the H atoms are randomly distributed within the matrix (outside the cavity) at all temperatures. Throughout the introduction of atoms and the dynamic evolution stage, the uncaptured H atoms demonstrate a tendency to migrate towards around the cavity, despite favoring the escape of interstitial H atoms with increasing temperature. As the process continues and further atoms are introduced, the cavity persists in trapping H atoms. At the end of the introduction of atoms (at 20 ns), the efficiency of H capture by the cavity was measured as $5.73 \times 10^8$ nm$^{-3}$s$^{-1}$ at 300 K, $11.33 \times 10^8$ nm$^{-3}$s$^{-1}$ at 723 K, and $8.03 \times 10^8$ nm$^{-3}$s$^{-1}$ at 973 K. During the dynamic evolution (after 20 ns), the $C_H$ inside the cavity remained nearly constant at 300 K. However, at higher temperatures, some H atoms within the cavities de-trapped from the surface and subsequently escape from the matrix, reducing $C_H$ both inside and outside the cavity. Despite this, the $C_H$ inside the cavity remained higher than in the surrounding matrix. The above results confirm that, even without continuous H introduction, it is undeniable that H adsorbed on the cavity surface under high-temperature conditions will gradually desorb, but the cavity surface has always been the most favorable position for H atoms during the desorption process [57].

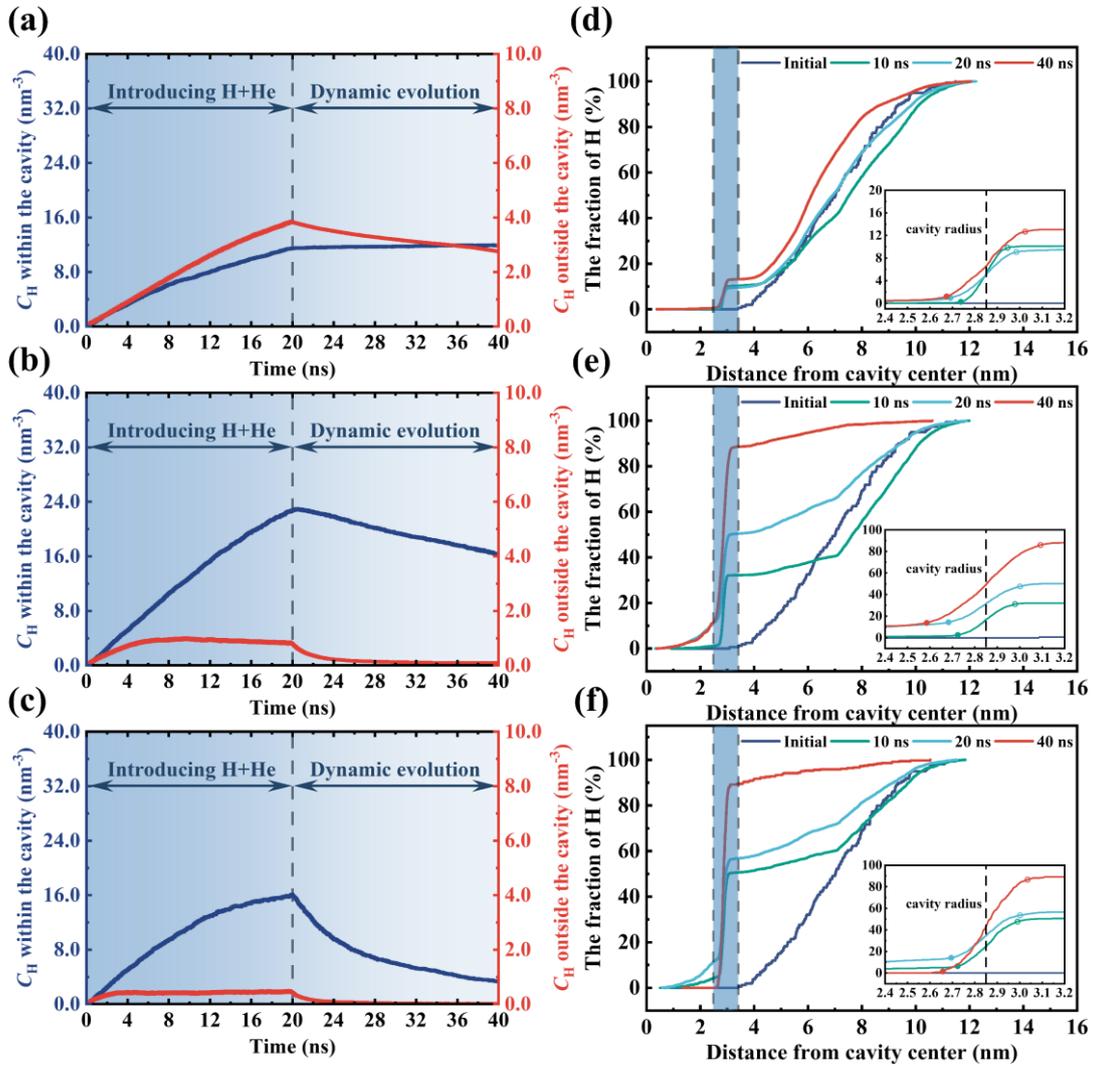

**Fig. 4.** (a-c) Evolution of H concentration ($C_H$) inside and outside the cavity at temperatures of (a) 300 K, (b) 723 K, and (c) 973 K during different stages. (d-f) The fraction of H within a distance of the cavity center at temperatures of (d) 300 K, (e) 723 K, and (f) 973 K, respectively. The local magnification map is represented by the blue banded region and serves to present the H fraction in the inner and outer surfaces of the cavity.

While the majority of H atoms remain trapped on both the inner and outer surfaces,

a non-negligible fraction diffuses into the cavity center, as seen in Figs. 4 (d) and (f). To further investigate the evolutionary behavior of the atoms inside the cavity, Fig. 5 conducts the radial PDF of H atoms at different instances in the entire system and the statistical analysis of the number of $H_2$ molecules at each temperature. Previous DFT and MC simulations [56,57] suggest that in an oversaturated state, the cavity surface becomes fully occupied by H atoms, enabling $H_2$ molecule formation in the cavity center if sufficient space is available. This is consistent with the H-H bond length of 0.74 Å observed in the radial PDF, matching the equilibrium bond length of $H_2$. These findings also align with the predictive model proposed by Hou et al. [61], which estimates that H atoms adsorbed on cavity surfaces form $H_2$ molecules in the core at a separation distance of ~ 0.74 Å. Correspondingly, the radial PDF shows a maximum H-H interaction probability at 2.51 ~ 2.57 Å at 300 K, indicating that most H atoms remain outside the cavity. However, at higher temperatures (723 K and 973 K), this peak shifts to 1.63 ~ 1.83 Å, confirming that trapped H atoms on the cavity surface reach the critical interaction distance for $H_2$ formation.

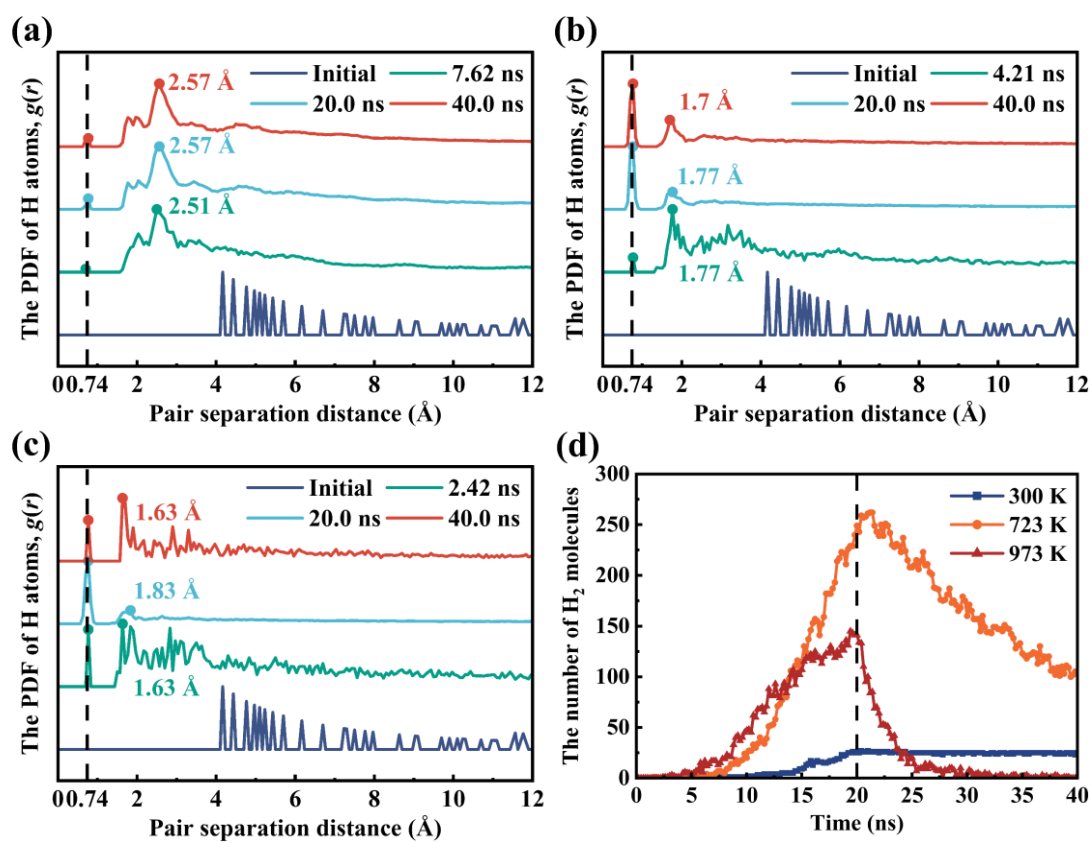

**Fig. 5.** (a-c) The radial pair distribution function (PDF) of H atoms at different

instances in the entire system at temperatures of (a) 300 K, (b) 723 K, and (c) 973 K, respectively. A pair separation distance of 0.74 Å in the radial PDF is indicative of the formation of $H_2$ molecules. (d) The statistical analysis of the number of $H_2$ molecules at each temperature.

The number of $H_2$ molecules formed exhibits a non-monotonic dependence on temperature, as presented in Fig. 5 (d). This is closely correlated with the temperature-dependent concentration of trapped H atoms inside the cavity Figs. 4 (a-c). At 973 K, $H_2$ nucleation occurs rapidly (within 2.42 ns), but the enhanced thermal mobility of interstitial H atoms simultaneously promotes their escape from the matrix, ultimately limiting the total trapped H content. In contrast, at 300 K, although H capture efficiency is reduced compared to higher temperatures, negligible H escape leads to highly stable configurations—with trapped H atoms on cavity surfaces and $H_2$ molecules at the center. Interestingly, the highest $C_H$ and largest number of $H_2$ molecules are observed at 723 K. It demonstrates that intermediate temperatures optimize two competing effects: (1) thermal activation of $H_2$ formation in the cavity center and (2) escape of interstitial H atoms. These results suggest the existence of an optimal H trapping temperature range within this broad range, where the trapping capacity of the cavity greatly exceeds the capacity for H escape.

To elucidate the atomic evolution within the cavity, Fig. 6 presents cross-sectional snapshots of the atomic configuration at key time points. These observations reveal a temperature-dependent competition between H trapping and escape mechanisms. As shown in Figs. 4-6, at 300 K, the cavity exhibits a stable H trapping regime where the interstitial H capture rate outweighs the escape rate over the entire simulation timeframe. At 723 K, the system is dominated by trapping behavior: H atoms adsorbed on the surface consistently form $H_2$ molecules in the cavity center during dynamic evolution. In contrast, at 973 K, the process undergoes a clear transition—beginning with $H_2$ formation in the center but ultimately shifting to a state where most H atoms remain on the cavity surface. This demonstrates a complete reversal from trapping-dominated to escape-dominated behavior. Furthermore, in all of the above cases, the cavity surface serves as the preferential trapping site for H atoms, leading to the formation of a "shell-like" structure within the cavity.

Additionally, two aspects are worth noting: First, as the cavity captures H atoms, He atoms are simultaneously trapped and remain stable at the cavity center. These He atoms occupy partial cavity space, which hinders $H_2$ molecule formation and thus affects H capture and retention. Section 3.2 provides a detailed analysis of how coexisting H and He atoms in BCC-Fe influence this process. Second, the ability of H atoms in the BCC-Fe matrix to be consistently trapped by the cavity is closely related to the $C_H$ in the matrix corresponding to different conditions (temperature, cavity size, and He/V ratio), which will be discussed in Section 4.2. The above results presuppose that the $C_H$ within the matrix satisfies the critical concentration at all three temperatures so that H atoms can be effectively trapped by the cavity.

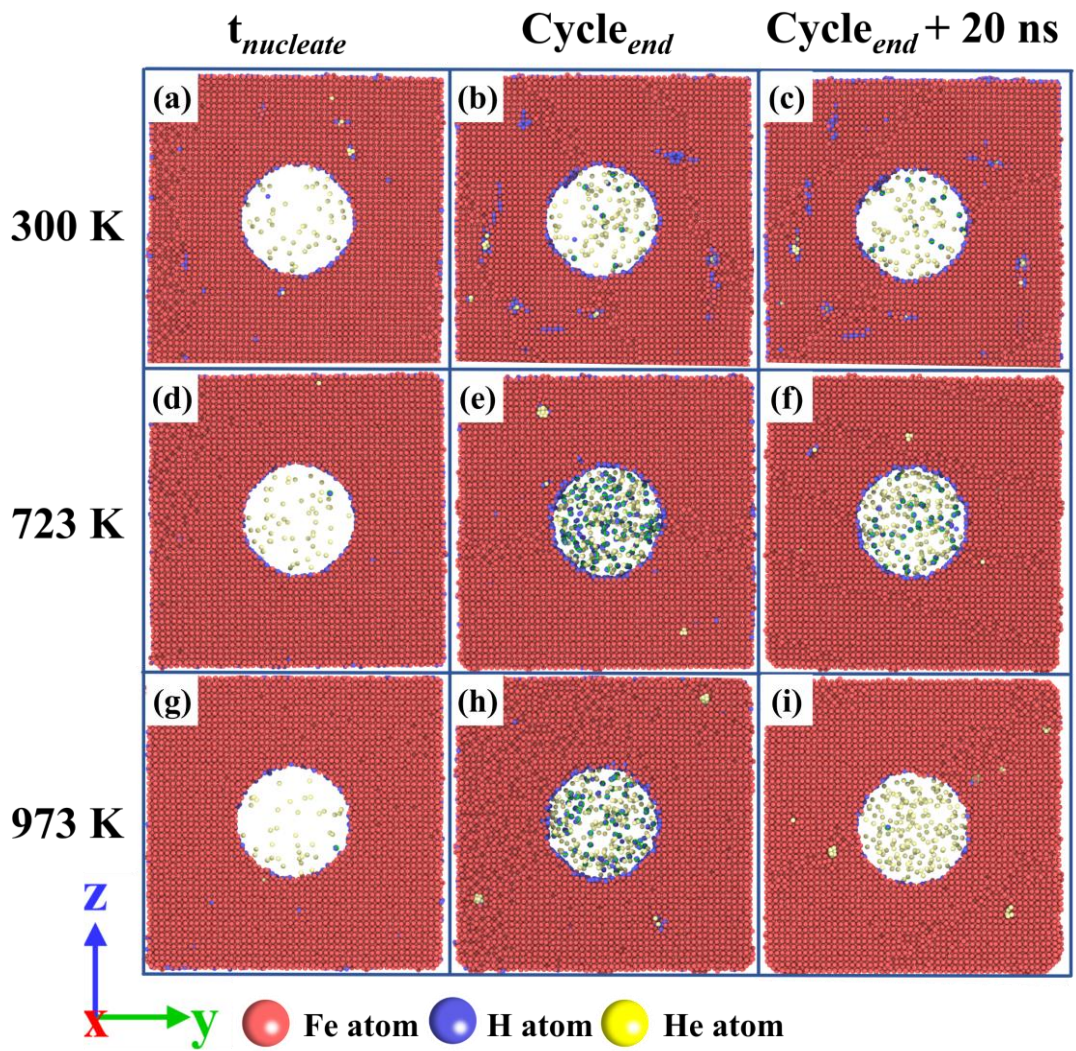

**Fig. 6.** Slice snapshots of atomic configurations at specified moments, where $t_{nucleate}$

corresponds to the formation of H₂ molecules. The snapshots are 10 Å in thickness to facilitate observation of the distribution of H/He within the cavity, and the bonds between H₂ molecules are colored green.

### *3.2 Cavity structure effects on the trapping of H*

Fig. 7 presents the evolution of H concentration ($C_H$) both inside and outside the cavity, the radial distribution of H atomic fractions, and the pair distribution function (PDF) of H atoms for cavities with different He/V ratios. To isolate and clearly analyze the effect of cavity structure on H trapping, only H atoms are introduced into the matrix at each cycle. Additionally, the kinetic evolution of H/He atom trapping in cavities of varying sizes is detailed in the Supplementary Materials. Notably, Figs. 7 (a) and (d) serve as a He-free control case that enables direct comparison with Figs. 4 (b) and (e), providing a systematic analysis of how He presence in the BCC-Fe matrix affects H trapping behavior under identical thermal conditions (723 K). The results indicate that interstitial He atoms facilitate the formation of $H_iHe_nV_m$ clusters in the matrix. Although these clusters exhibit lower H-binding energies than cavities, their proliferation creates numerous alternative H-trapping sites. This introduces competitive trapping effects that reduce H accumulation at pre-existing cavities.

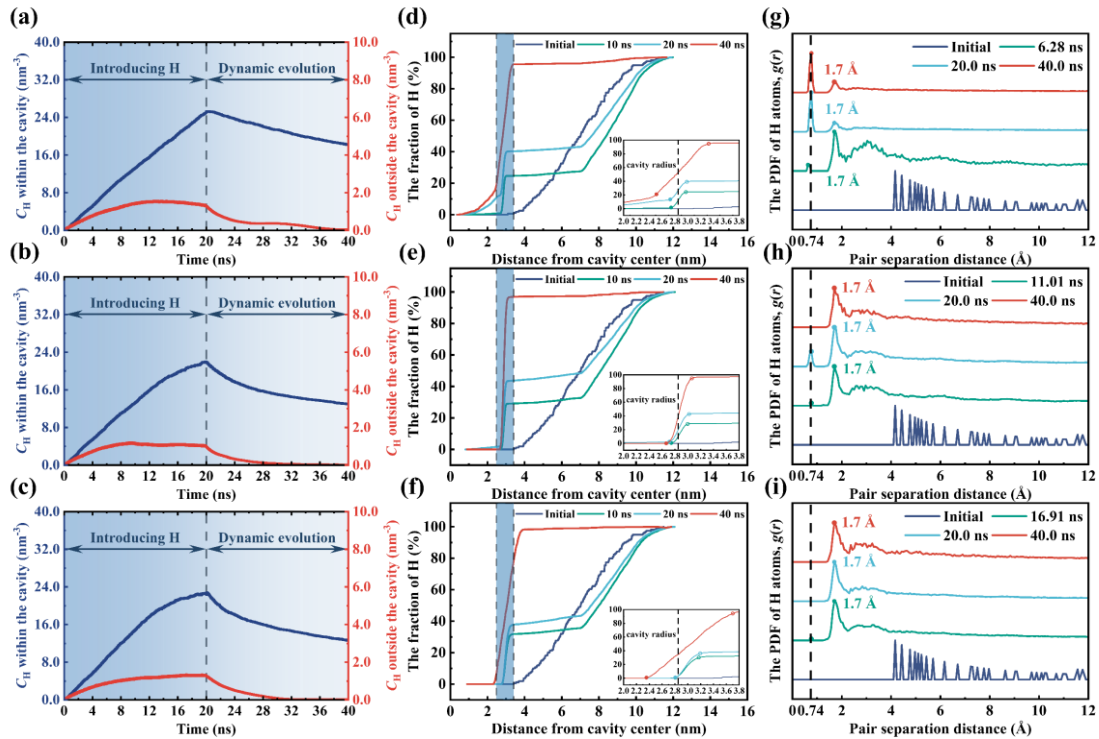

**Fig. 7.** (a-c) Evolution of H concentration ($C_H$) inside and outside the cavity at 723 K in different stages with varying He/V ratios of (a) 0, (b) 0.5, and (c) 0.8, respectively. (d-f) The fraction of H atoms within a distance of the cavity center corresponds to different He/V ratios. The local magnification map is represented by the blue banded region and serves to present the H fraction in the inner and outer surfaces of the cavity. (g-i) The radial pair distribution function (PDF) of H atoms corresponding to different He/V ratios. A pair separation distance of 0.74 Å in the radial PDF is indicative of the formation of $H_2$ molecules.

As demonstrated in Figs. 7 (a-c), He-containing cavities exhibit reduced H trapping capability compared to pure voids (He/V=0). At the end of the introduction of atoms (at 20 ns), the efficiency of H capture by the cavity was measured as $12.55 \times 10^8$ $nm^{-3}s^{-1}$, $10.91 \times 10^8$ $nm^{-3}s^{-1}$, $11.28 \times 10^8$ $nm^{-3}s^{-1}$ for the three cavity ratios, respectively. Furthermore, as shown in Figs. 7 (d-f), the He/V ratios of 0, 0.5, and 0.8 correspond to HJ fractions within the cavity of 40.3%, 31.62%, and 34.93%, respectively. During the dynamic evolution (after 20 ns), a fraction of the trapped H is de-trapped from the cavity surface and subsequently escapes from the substrate. The $C_H$ corresponding to the three He/V ratios decreased to 18.25 $nm^{-3}$, 12.97 $nm^{-3}$, and 12.68 $nm^{-3}$ at 40 ns, respectively. Nevertheless, the $C_H$ profiles reveal significantly higher values within cavities than in the surrounding matrix for all investigated He/V ratios, indicating preferential H accumulation at these defect sites despite the presence of He atoms. Thus, it can be hypothesized that during irradiation without continuous H introduction, the limited H inventory within the matrix should become preferentially trapped at cavity surfaces. Consequently, when the H-He-V cavity is analyzed experimentally by EELS after irradiation [10], the high probability of detecting a peak corresponding to H should correspond to the cavity surface.

Furthermore, most of the trapped H atoms enter the cavity center to form $H_2$ molecules, which are also present in the cavity during the dynamic evolution, as depicted in Fig. 7 (g). In contrast, at high He/V ratios, only a few $H_2$ molecules form, even when the H-H interaction distance on the cavity surface approaches the critical value of ~1.7 Å [57], as evidenced by Figs. 7 (h) and (i). At a He/V ratio of 0.8, $H_2$ production is notably suppressed during the introduction of atoms stage, as shown in Fig. 8 (c). Later, the H atoms primarily de-trap and escape from the cavity surface, further inhibiting $H_2$ formation within the cavity. The aforementioned results are attributed to two aspects: Firstly, the electron density near the cavity surface increases with the increase of He content in the cavity, thus unfavorable for the H atoms in the matrix to be trapped by the cavity surface [25,62]. Secondly, the presence of He within cavities creates steric constraints that prevent sufficient free volume for stable $H_2$ formation [57], thereby reducing H retention capacity.

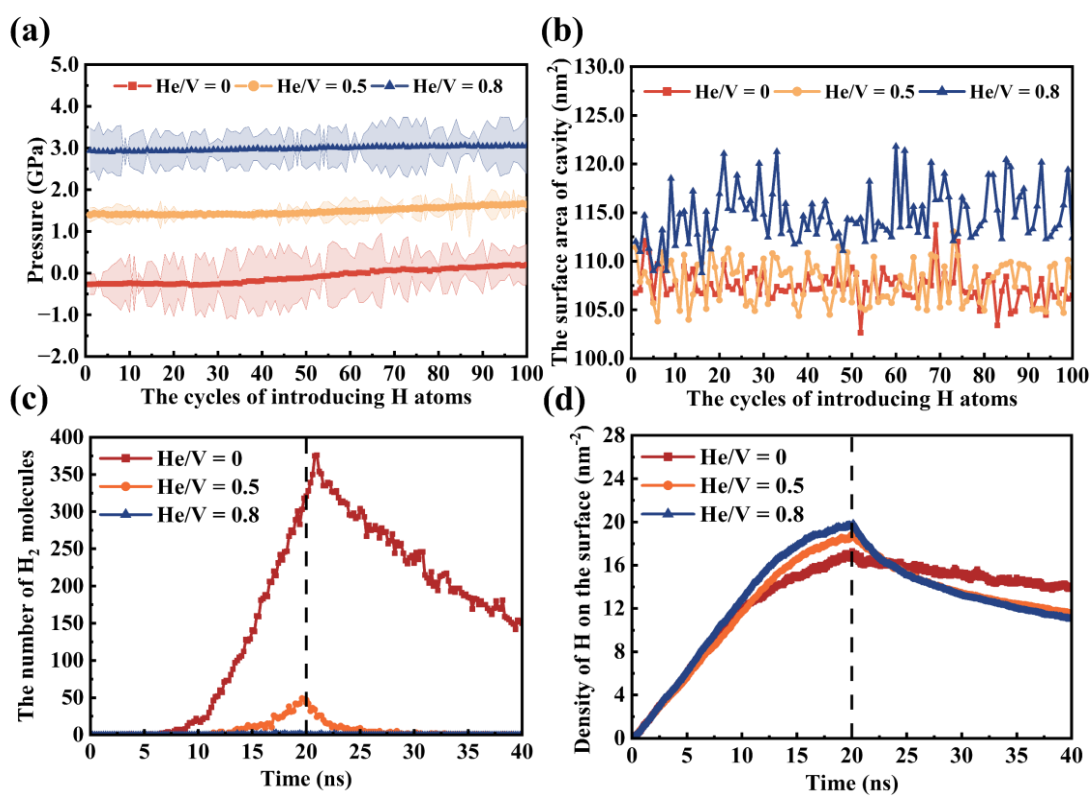

**Fig. 8.** (a) The evaluation of the pressure inside the cavity at each He/V ratio. (b) The evolution of the corresponding surface area of the cavity with three He/V ratios. (c) The statistical analysis of the number of $H_2$ molecules at each cavity structure. (d) The density of H atoms trapped on the inner and outer surfaces of the cavity for all He/V ratio conditions.

To elucidate the behavior of H atoms trapped by cavities with high He/V ratios, Fig. 8 presents the pressure, the number of $H_2$ molecules inside the cavity, the surface area of the cavity, and the density of H atoms on both the inner and outer surfaces at each He/V ratio. The pressure generated by $H_2$ molecules or He in the cavity at each He/V ratio is shown in Fig. 8 (a). At He/V = 0, the production of $H_2$ molecules in the cavity is enhanced as the cavity continues to trap H atoms, resulting in an increase in the intracavity pressure from the initial value of -0.27 to 0.21 GPa. At He/V = 0.5, the intracavity pressure increases from the initial 1.41 to 1.65 GPa, due to the relatively small number of $H_2$ molecules formed in the cavity. At He/V = 0.8, since almost no $H_2$ molecules are formed in the cavity center, the corresponding intracavity pressure is contributed only by He and remains around 3.0 GPa across the entire simulated progress.

The increased pressure within the cavity further modifies the surface area, as shown in Fig. 8 (b). Consequently, as the He/V ratio increases, the expanding surface creates additional sites for H atom adsorption. Except for the $H_2$ molecules that have been formed within the cavity, the corresponding H atoms trapped by the cavity surface increase with the increase of the He/V ratio, as evidenced by Fig. 8 (c) and (d). Moreover, Fig. 9 presents snapshots of slices of the Von Mises shear strain distributions at various He/V ratios, where the strain field regions of the different cavities at specific moments are highlighted by dashed circles. As the He/V ratio increases, the Von Mises shear strain induced by He atoms surrounding the cavity grows proportionally, accompanied by an expansion of the corresponding atomic strain field in the BCC-Fe matrix. This creates favorable conditions for H accumulation, as H atoms are attracted to both the surface of the cavity and the stress field generated by the pressurized cavity [23].

Combined with the results of Figs. 7 (d-f), the H fraction within the distance from the cavity center corresponding to different He/V ratios allows a preliminary assessment of the capture distance of different cavity structures. The capture distance corresponding to the three He/V ratios are approximately 0.3148 nm, 0.3205 nm, and 0.5178 nm, respectively. This demonstrates that the capture distance of H in the cavity increases with the increase of the He/V ratio in the range of 0 ~ 0.8. Unlike cavities with low He/V ratios (including voids), those with high He/V ratios demonstrate distinct H-trapping behavior. The elevated internal pressure from accumulated He expands the surface area of the cavity while simultaneously increasing the effective trapping distance, resulting in greater H capture capacity.

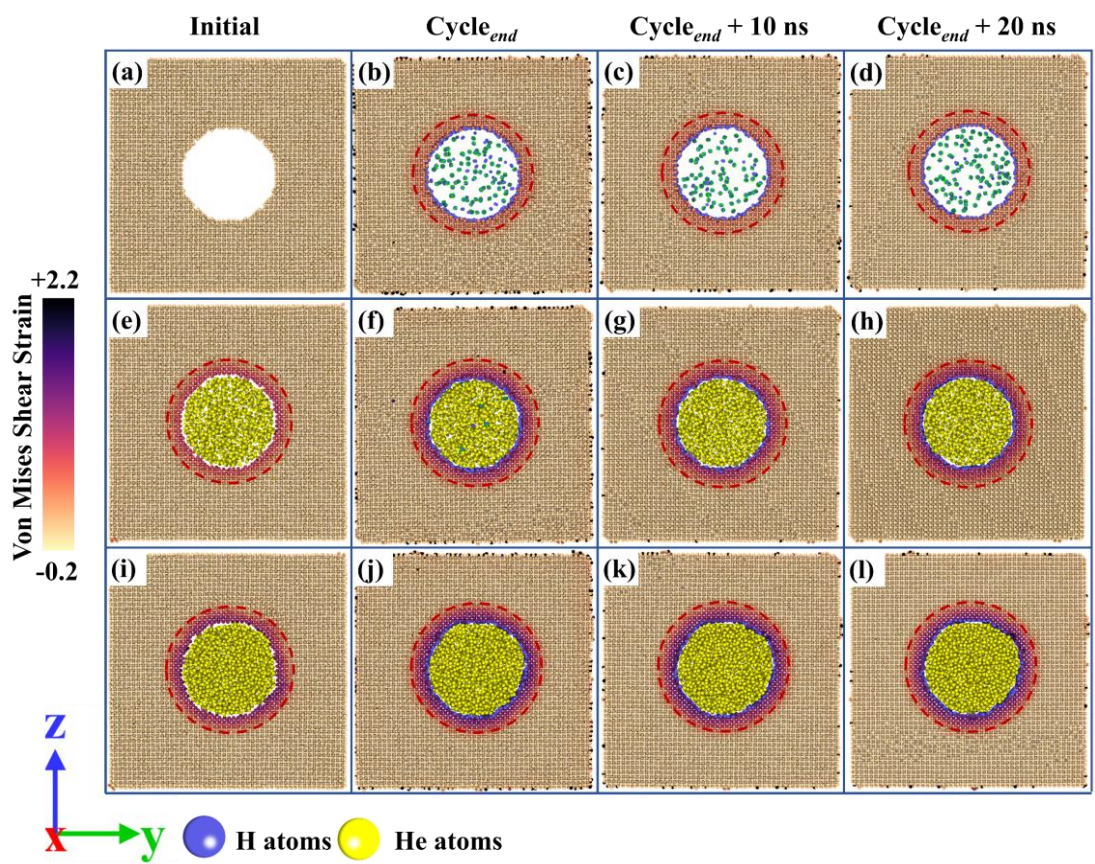

**Fig. 9.** Slice snapshots of the Von Mises shear strain distribution at 723 K with He/V ratios of (a-d) 0, (e-h) 0.5 and (i-l) 0.8, respectively. The regions of the strain field for different He/V cavities at specified moments are highlighted by dashed circles. The snapshots are 10 Å in thickness to facilitate observation of the distribution of H/He within the cavity, and the bonds between $H_2$ molecules are colored green.

## 4. Discussion

The recent experimental findings [10,18] offer valuable insights into the interaction of H and He atoms with cavities, as well as the formation of H-He-V cavity configurations following the trapping of H and He by cavities in various RAFM steels. In the aforementioned simulations, the current MD computational model references the pivotal experimental configurations outlined in the recent study by G.was's research group [10]. This includes the specified irradiation temperatures, H/He introduction ratios, and the root mean cubed cavity diameter formed at the corresponding dpa. Although the rates of H and He introduction in the current MD model are considerably higher than those in the experiments, the concentration of H/He atoms introduced by the matrix is consistent with the experimental conditions. Many recent studies employing similar approaches have demonstrated good agreement with experiments [63,64]. From a qualitative perspective, at the spatial and temporal scales of MD simulations, this provides a novel insight into the evolutionary behavior between cavities and H/He atoms. Additionally, the thermodynamic analyses also provide substantial evidence in support of the formation of core-shell structures by H-He-V cavities in the RAFM steel irradiation experiments.

### *4.1 Kinetic evolution of H-He coexisting cavities*

When H and He atoms coexist in the BCC-Fe matrix, the kinetic evolution of He trapping by cavities is illustrated in Figs. 2 and 3. These results demonstrate that the increasing temperature elevates the kinetic energy of He atoms, thereby effectively suppressing the self-trapping effect of He atoms and destabilizing the formation of $H_iHe_nV_m$ clusters. Consequently, interstitial He atoms exhibit higher probabilities of either escaping the lattice or being trapped by cavities, with elevated temperatures promoting cavity-mediated He retention. Moreover, the number of He atoms trapped by cavities increases with cavity size (0.57 nm, 1.43 nm, and 2.86 nm) at a fixed temperature of 723 K, as shown in Figs. S1. Conversely, the size and number of $He_nV_m$ clusters formed by He self-trapping within the matrix (outside cavities) decrease as cavity size increases. Furthermore, the results presented in Fig. 3 demonstrate that the presence of interstitial H atoms impedes the migration of interstitial He atoms, thereby promoting the self-trapping of $He_nV_m$ clusters. The trapping of H/He atoms by these clusters results in the formation of more $H_iHe_nV_m$ clusters, and the H and He atoms in the matrix are not favored for trapping by the pre-existing cavity. Notably, under varying temperatures and cavity structures, and irrespective of H presence in the BCC-Fe matrix, He atoms trapped by cavities remain stably positioned at their centers, forming a persistent "core-like" structure.

Importantly, the irradiation temperature as well as the cavity structure (including the size and He/V ratio) have a significant effect on the trapping of H atoms by the cavity. The increase in temperature, although leading to the escape of interstitial H atoms from the matrix, also promotes the formation of $H_2$ molecules in the cavity to store more H, as presented in Figs. 4-6. This behavior demonstrates a temperature-dependent competition between H-atom escape and trapping mechanisms. A comparison of Figs. 3 and 8 reveal that interstitial He atoms promote the formation of $H_iHe_nV_m$ clusters in the matrix. While these clusters exhibit lower H-binding energies compared to cavities, they create numerous additional trapping sites beyond the primary cavity regions. This competitive trapping mechanism effectively redistributes H atoms, reducing their accumulation at pre-existing cavities. Moreover, as shown in Figs. 7-9, the available space for $H_2$ formation within cavities decreases with increasing He/V ratio, which limits H storage capacity. However, the pressure exerted by trapped He atoms expands the cavity surface area, partially compensating for this effect by creating additional trap sites. Furthermore, at high He/V ratios, the strain field generated by He atoms extends beyond the cavity, effectively broadening the trapping region. In all studied cases, the cavity surface acts as the energetically favorable trapping site for H atoms, resulting in a stable "shell-like" structure within the cavity.

Although the direct H-He pair interaction is repulsive in both vacuum and the BCC-Fe lattice [22,25], their mutual attraction to low-electron-density regions (e.g., vacancies) enables coexistence in cavities. The formation of He-V and H-V binding forces effectively screens the repulsive interaction between trapped He and H atoms [19,22,25], allowing their stable co-occupation of vacancy-type defects. This screening effect exhibits a strong dependence on the He content and size of the cavity, with mitigation of H-He repulsion observed in larger cavities and at lower He/V ratios, as shown in Figs. 7 and S4. Under various conditions, the kinetic evolution results demonstrate that both H and He in the matrix are simultaneously trapped by the cavity until reaching saturation. The trapped H and He atoms occupy distinct energetically favorable sites within the cavity, resulting in a core-shell structure of the H-He-V cavity.

### *4.2 Thermodynamic analysis of H-He coexisting cavities*

The de-trapping of H from the cavity surface will occur at about 327 °C (600 K)

[21], it is generally accepted that above this de-trapping temperature, cavities cannot effectively trap H atoms from the BCC-Fe matrix. However, Clowers et al. [10] provided clear evidence for the H-He-V cavity configuration at an irradiation temperature of at irradiation temperature 450 °C (723 K) through the use of EELS elemental mapping. This analysis revealed that H forms a halo structure at the periphery of the cavity, while He is present inside the cavity. Similarly, the results of the present MD simulations demonstrate that, under these conditions, the majority of H and He atoms are found at the surface and center of the cavity, respectively, and the H-He-V cavities exhibit a similar core-shell structure. Although the agreement between the current MD computations and experimental observations confirms the validity and accuracy of the simulation results, further thermodynamic studies are required to fully assess the feasibility of cavity trapping of H atoms at different temperatures as well as H concentrations. The alteration in the Gibbs free energy of the system ($\Delta G$) during the process of trapping the H atoms within the cavity is expressed as:

$$\Delta G = \Delta H - T\Delta S \qquad (4)$$

where $\Delta H$ and $\Delta S$ are the enthalpy and entropy changes, respectively, associated with the trapping process. The $T$ is the thermodynamic temperature. According to the accepted definition [59,62], the enthalpy change can be determined as follows:

$$\Delta H = -\sum_{i=2}^{z} E_b^v(0,0,i) - \sum_{j=1}^{y} E_b^{He}(x,j,z) - \sum_{l=1}^{x} E_b^H(l,y,z) \qquad (5)$$

where the $x$, $y$, and $z$ denote individual H atoms, He atoms, and vacancies, respectively. The terms in the equation represent the binding energy formed between vacancies, the binding energy of He atoms to vacancies, and the binding energy of H atoms to vacancies, respectively.

The entropy change can be approximately given by:

$$\Delta S = -(z-1)k_B \ln \frac{C_V}{1-C_V} - yk_B \ln \frac{C_{He}}{1-C_{He}} - xk_B \ln \frac{C_H}{1-C_H} \qquad (6)$$

where $k_B$ is the Boltzmann constant, $C_V$, $C_{He}$ and $C_H$ are the bulk atomic concentration of vacancy, and He, H in BCC-Fe (i.e. V/Fe, He/Fe, and H/Fe ratios).
Based on Eqs. (4-7), the Gibbs free energy change of the system is defined as [62]:

$$\Delta G(x,y,z) = -\left\{\sum_{i=2}^{z} E_b^v(0,0,i) + \sum_{j=1}^{y} E_b^{He}(x,j,z) + \sum_{l=1}^{x} E_b^H(l,y,z)\right\} -$$

$$\left\{(z-1)k_\text{B}T\ln\frac{C_\text{V}}{1-C_\text{V}} + yk_\text{B}T\ln\frac{C_\text{He}}{1-C_\text{He}} + xk_\text{B}T\ln\frac{C_\text{H}}{1-C_\text{H}}\right\} \quad (7)$$

As a stable cavity (void or bubble) has been constructed in the initial system, this study is mainly concerned with the process of H atoms being trapped by the cavity. Consequently, the change in the Gibbs free energy of the system caused by the $\text{He}_n\text{V}_m$ and vacancy clusters can be neglected. Therefore, when the initially constructed cavity in the system is a stable structure (constant He/V ratio), Eq. (7) can be simplified as follows:

$$\Delta G(x,y,z) = -\sum_{l=1}^{x} E_\text{b}^\text{H}(l,y,z) - xk_\text{B}T\ln\frac{C_\text{H}}{1-C_\text{H}} \quad (8)$$

If the reversible work corresponding to the trapping of H atoms by a cavity at different temperatures and H concentrations is negative, i.e., $\Delta G$ (x+1, y, z) - $\Delta G$ (x, y, z) < 0, this suggests that the trapping of H atoms by the cavity proceeds spontaneously. When the H concentration within the bulk and cavity structure is given, the minimum values ($\Delta G_\text{min}$) corresponding to the Gibbs free energy in cavities can be calculated. It is worth noting that the adopted $E_\text{b}^\text{H}(l,y,z)$ energetics parameters are given by our previous studies [25,62].

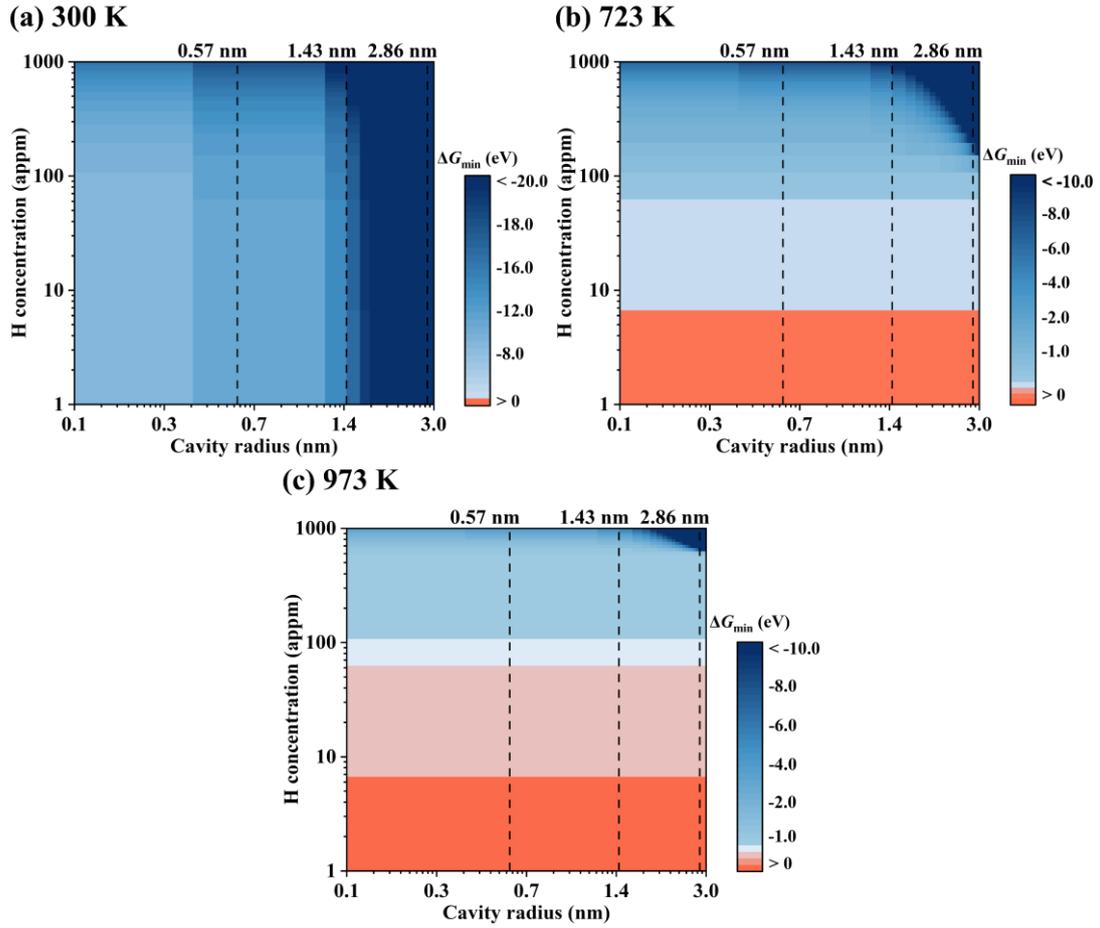

**Fig. 10.** Statistical thermodynamic analysis of H atoms trapped by cavities (He/V = 0) at different temperatures. The minimum Gibbs free energies ($\Delta G_{min}$) of the systems are determined for varying H concentrations in the matrix and different cavity sizes. The three reference cases in the figure correspond to cavities with radii of 0.57 nm, 1.43 nm, and 2.86 nm in the aforementioned MD simulation models.

Fig. 10 illustrates the evolution of the $\Delta G_{min}$ for cavity-captured H under different H concentrations (1-1000 appm) and different cavity sizes (radius of ~ 0.1-3 nm) within the BCC-Fe matrix. Even at high temperatures, $\Delta G_{min} < 0$ occurs if the H concentration in the matrix is sufficiently high. It is thermodynamically favorable for H atoms to be trapped by the cavity, provided the H concentration exceeds a critical threshold. In the case of 300 K, the results demonstrate that even if the H concentration is particularly low (~ 1 appm), the H atoms are still trapped by each cavity under concentration-driven and thermodynamic effects. Figs. 10 (b) and (c) demonstrate that H trapping remains thermodynamically favorable ($\Delta G_{min} < 0$) when the H concentration exceeds a critical value, even at temperatures above the desorption threshold (600 K). Specifically, critical H concentrations of approximately 10 appm and 100 appm correspond to temperatures of 723 K and 973 K, respectively. Besides, combined with the previous study [62], thermodynamic calculations confirm that He being trapped in the cavity to form a "core-like" structure is energetically more favorable than H being trapped on the surface to form a "shell-like" structure.

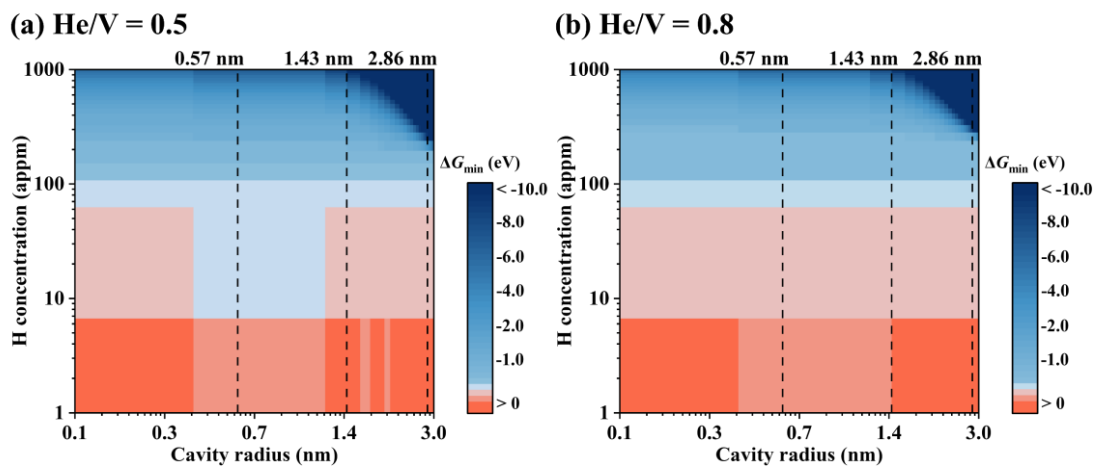

**Fig. 11.** Statistical thermodynamic analysis of H atoms trapped by cavities in different He/V ratios at 723 K. The minimum Gibbs free energies ($\Delta G_{min}$) of the systems are determined for varying H concentrations in the matrix and different cavity sizes.

Fig. 11 presents the statistical thermodynamic analysis of H atoms trapped in various sizes of cavities with the He/V ratios of 0.5 and 0.8 at 723 K. Comparison with Fig. 10 (b) shows that at constant temperature, the critical H concentration required for cavity trapping increases with higher He/V ratios, demonstrating a repulsive interaction between H and He within the cavity. These results agree with DFT calculations [22,25], which indicate that increasing He content within the cavity enhances electron density near the cavity surface, thereby reducing its H trapping capability. Furthermore, under the same conditions, it is evident that the larger the cavity size, the easier it is to trap H atoms. It should be noted that the current thermodynamic analysis is limited to a perfect BCC-Fe lattice containing initial cavities. The value of $\Delta G_{min}$ corresponds to the most probable value of H trapping by the cavity, which is determined by the actual defects-to-H action variations within the matrix.

In the recent irradiation experiments [10,11], the damage level was up to 50 dpa and the H/He implantation rate was up to 40/10 appm/dpa. The irradiation process is accompanied by a high damage level and a continuous accumulation of H concentration. Accordingly, this suggests that under these experimental conditions, even at high temperatures and high He/V ratios, the H concentration in the BCC-Fe matrix readily reaches the critical value required for cavity trapping. In addition, quantitative methods for predicting the energy of He-cavity interactions are presented in our previous study [62]. The results demonstrate that it is thermodynamically more favorable for He atoms to be captured by the cavity under the same conditions as compared to H. It is important to note that, despite a slight reduction in the binding energy of the H atoms to the bubble surface when the cavity is a He bubble [23], the qualitative results of the aforementioned thermodynamic analyses remain unaltered. Therefore, the above results of the kinetic evolution of H and He atoms coexisting in the BCC-Fe matrix with cavity (including the core-shell structure formed by the H-He-V cavity) are reliable and accurate as long as the H/He concentration in the matrix reaches a critical concentration at the corresponding temperature and He/V ratio.

## 5. Conclusion

From a kinetic perspective, the MD simulation results provide a detailed dynamic evolution of the H and He atoms coexisting in the BCC-Fe matrix interacting with the cavity. When H and He atoms coexist in the matrix, higher temperatures suppress He self-trapping, promoting He escape or cavity trapping. Within the studied temperature range (300 ~ 973 K), cavity-mediated He trapping becomes the dominant mechanism at elevated temperatures. However, the presence of interstitial H atoms impedes migration, enhancing additional $H_iHe_nV_m$ cluster formation in the matrix, thereby reducing the accumulation of H/He in the pre-existing cavity. Besides, elevated temperatures drive interstitial H escape from the matrix but enhance $H_2$ molecule formation in cavities, reflecting competition between H escape and cavity-mediated storage. Furthermore, higher He/V ratios reduce cavity space for $H_2$ molecules, limiting H storage. This is partially offset by He-induced cavity expansion, which increases surface trap sites. At high He/V ratios, strain fields from trapped He atoms extend beyond the cavity, broadening the effective H-trapping region.

From a thermodynamic perspective, the trapping of H and He atoms by cavities becomes increasingly unfavorable with rising temperature, which consequently requires higher critical H/He concentrations in the matrix to maintain stability. Furthermore, the trapping of H atoms by cavities is also thermodynamically disfavored as the He/V ratio increases. Nevertheless, the kinetic evolution results for H and He co-existing in the BCC-Fe matrix with cavity are reliable and accurate when the matrix solute concentrations reach the critical threshold under the given conditions. Consequently, the H-He-V cavity stabilizes into a core-shell structure, with H and He atoms preferentially occupying distinct energetically favorable sites within the cavity. Moreover, the trapping of He by the cavity to form a "core-like" structure is thermodynamically more favorable than the trapping of H by the cavity to form a "shell-like" structure. Critically, thermodynamic calculations align with kinetic evolution results, providing substantial support for the formation of the core-shell structure of the H-He-V cavities in the high-temperature irradiated RAFM steel experiments.

## Declaration of competing interest

The authors declare that they have no known competing financial interests or personal relationships that could have appeared to influence the work reported in this paper.

## Acknowledgments

This work was supported by the National MCF Energy R&D Program 2022YFE03110000; the National Natural Science Foundation of China (Grant Nos. 12422514, 12192280, 11935004, and 12275009); the National Key R&D Program of China under Grant No. 2025YFB3003600. The authors gratefully acknowledge the computing resources and support provided by the High-Performance Computing Platform of the Center for Life Science (Peking University).